\documentclass[twocolumn, switch]{article} 
\usepackage{preprint}

\usepackage{amsmath, amsthm, amssymb, amsfonts}
\usepackage[numbers,sort,square]{natbib}

\usepackage[utf8]{inputenc}	
\usepackage[T1]{fontenc}	
\usepackage{xcolor}		
\usepackage[hidelinks]{hyperref}	
\usepackage{booktabs} 		
\usepackage{nicefrac}		
\usepackage{microtype}		
\usepackage{lineno}		
\usepackage{float}

\usepackage{fancyhdr}
\usepackage[nolist,nohyperlinks]{acronym}               
\usepackage{pifont} 
\usepackage[linesnumbered,ruled,vlined]{algorithm2e}    
\usepackage{algorithmic}								
\usepackage{enumitem}									
\usepackage{url}                                        
                               
\DeclareMathAlphabet{\mathcal}{OMS}{cmsy}{m}{n}			

\newcommand{\around}{\raise.17ex\hbox{$\scriptstyle\sim$}}

\setlist[itemize,enumerate]{leftmargin=1.8em, itemindent=0em}

\begin{acronym}
	\acro{AI}{artificial intelligence} 
	\acro{CLI}{command-line interface}
	\acro{GUI}{graphical user interface}
	\acro{HMAC}{hash-based message authentication code}
	\acro{JVM}{Java virtual machine}
	\acro{LLM}{large language model}
	\acro{NLP}{natural language processing}
	\acro{OCR}{optical character recognition}
	\acro{OTP}{one-time pad}
	\acro{SNOW}{Steganographic Nature of Whitespace}
\end{acronym}

\usepackage{newfloat}
\DeclareFloatingEnvironment[name={Supplementary Figure}]{suppfigure}
\usepackage{sidecap}
\sidecaptionvpos{figure}{c}

\usepackage{titlesec}
\titlespacing\section{0pt}{12pt plus 3pt minus 3pt}{1pt plus 1pt minus 1pt}
\titlespacing\subsection{0pt}{10pt plus 3pt minus 3pt}{1pt plus 1pt minus 1pt}
\titlespacing\subsubsection{0pt}{8pt plus 3pt minus 3pt}{1pt plus 1pt minus 1pt}

\usepackage{tikz,xcolor,hyperref}

\definecolor{lime}{HTML}{A6CE39}
\DeclareRobustCommand{\orcidicon}{
	\begin{tikzpicture}
		\draw[lime, fill=lime] (0,0) 
		circle [radius=0.16] 
		node[white] {{\fontfamily{qag}\selectfont \tiny ID}};
		\draw[white, fill=white] (-0.0625,0.095) 
		circle [radius=0.007];
	\end{tikzpicture}
	\hspace{-2mm}
}
\foreach \x in {A, ..., Z}{\expandafter\xdef\csname orcid\x\endcsname{\noexpand\href{https://orcid.org/\csname orcidauthor\x\endcsname}
		{\noexpand\orcidicon}}
}

\title{Innamark: A Whitespace Replacement Information-Hiding Method}

\newcommand{\Footerhint}{\centering\footnotesize\color{gray!90}This is the author's version of a paper, published in the \emph{IEEE Access} journal under the \href{https://creativecommons.org/licenses/by/4.0/}{CC BY 4.0} license. See \href{https://doi.org/10.1109/ACCESS.2025.3583591}{doi:10.1109/ACCESS.2025.3583591}.}

\fancypagestyle{firstpage}{
	\fancyhf{}
	\fancyfoot[RE,LO]{\Footerhint}
}

\AtBeginDocument{\setlength{\footskip}{24pt}}

\usepackage{authblk}

\author[1]{Malte Hellmeier\orcidA{}}
\author[2]{Hendrik Norkowski\orcidB{}}
\author[1]{Ernst-Christoph Schrewe\orcidC{}}
\author[1]{\\Haydar Qarawlus\orcidD{}}
\author[3,1]{Falk Howar\orcidE{}}

\affil[1]{Fraunhofer ISST, 44147 Dortmund, Germany}
\affil[2]{Montsecure GmbH, 44799 Bochum, Germany}
\affil[3]{TU Dortmund, 44227 Dortmund, Germany}

\begin{document}
	
	\twocolumn[ 
	\begin{@twocolumnfalse} 
		
		\maketitle
		\thispagestyle{firstpage}
		
		\begin{abstract}
			Large language models (LLMs) have gained significant popularity in recent years. Differentiating between a text written by a human and one generated by an LLM has become almost impossible. Information-hiding techniques such as digital watermarking or steganography can help by embedding information inside text in a form that is unlikely to be noticed. However, existing techniques, such as linguistic-based or format-based methods, change the semantics or cannot be applied to pure, unformatted text.
			In this paper, we introduce a novel method for information hiding called Innamark, which can conceal any byte-encoded sequence within a sufficiently long cover text. This method is implemented as a multi-platform library using the Kotlin programming language, which is accompanied by a command-line tool and a web interface. By substituting conventional whitespace characters with visually similar Unicode whitespace characters, our proposed scheme preserves the semantics of the cover text without changing the number of characters. Furthermore, we propose a specified structure for secret messages that enables configurable compression, encryption, hashing, and error correction. 
			An experimental benchmark comparison on a dataset of 1\,000\,000 Wikipedia articles compares ten algorithms. The results demonstrate the robustness of our proposed Innamark method in various applications and the imperceptibility of its watermarks to humans. We discuss the limits to the embedding capacity and robustness of the algorithm and how these could be addressed in future work.
		\end{abstract}
		\keywords{Blind watermarking \and copyright protection \and data hiding \and data sovereignty \and digital text watermarking \and information hiding \and steganography \and Unicode characters.} 
		\vspace{0.73cm}
		
	\end{@twocolumnfalse} 
	]

	\section{Introduction}
	\label{sec:introduction}
	Interest in \acp{LLM} has grown rapidly in recent years, with a variety of promising applications emerging for individuals and businesses. The enduring process of digitization has led to numerous documents being stored directly in a machine-readable format that can be processed with \acp{LLM}. However, it is becoming increasingly difficult to protect this intellectual property, especially when data are shared. In addition, recent improvements in AI-generated text make it increasingly challenging to distinguish between text written by a human and text generated by an \ac{LLM}~\cite{Christ.2024}. This can make it more complicated to maintain control over data, often discussed under the umbrella term \emph{data sovereignty}~\cite{Jarke.2019,Hellmeier.2023,Scherenberg.2024}. Information-hiding techniques such as digital watermarking or steganography can help to address these concerns by hiding information in a cover document~\cite{Petitcolas.1999,Por.2012,Alkawaz.2016,Majeed.2021,Ahvanooey.2022}. 
	
	Researchers and practitioners have developed a variety of methods for hiding data in different files, such as image, video, audio, or text files, of which the latter are the most challenging~\cite{Alkawaz.2016,Bertini.2019}. Those techniques can be classified as \emph{watermarking} or fingerprinting methods, which focus on robustly storing copyright information and securing intellectual property, and \emph{steganography} methods, which focus on encoding secret information imperceptibly~\cite{Petitcolas.1999,Por.2012,Alkawaz.2016,Rizzo.2016}. To hide information in text, format-based methods shift words or lines or change fonts or text colors, whereas linguistic-based methods replace words with synonyms or generate new text~\cite{Majeed.2021,Zhang.2024}. Recently, researchers have proposed and discussed the possibilities for watermarking and steganography in \acp{LLM}~\cite{Kirchenbauer.2023,Christ.2024b,Steinebach.2024,Dathathri.2024,Xu.2024}. This is motivated by political regulations like the European Union's AI Act, which states that AI-generated texts must be detectable and mentions watermarks and fingerprints as possible solutions~\cite{EuropeanCommission.2024}.
	
	However, existing approaches show significant drawbacks in real-world use cases. For example, words cannot be replaced with synonyms in legal cover texts, as small semantic changes can have profound legal impacts. Format-based methods that employ line shifts or font colors cannot be applied to plain text files. Other methods using specific characters like Unicode confusables or spaces without width, called zero-width characters, are either visually recognizable by humans or not robust in different applications when using simple copy-and-paste tampering attacks~\cite{Ahvanooey.2022}.
	
	To close this gap, we extend the method proposed in~\cite{Hellmeier.2025} and call it Innamark. This method can hide any byte-encoded string inside a cover text by replacing all classical whitespace characters with a curated set of similar-looking whitespace characters from the Unicode standard~\cite{TheUnicodeConsortium.2025}. This leads to a text with a secret message hidden inconspicuously inside. This approach is applicable to plain text in different file formats due to its format independence, and the secret message remains when copied and pasted without alerting users. Our design can support optional functionalities like compression, encryption, hashing, or error-correcting codes. This makes it usable for both watermarking and steganography use cases. To sum up, the paper makes the following concrete contributions:
	\begin{itemize}
		\item We propose Innamark, an information-hiding technique that can hide any byte-encoded sequence inside a cover text.
		\item We provide a Kotlin library as a reference implementation.
		\item We introduce InnamarkTags, a specific secret message structure, to enable compression, encryption, hashing, and error-correcting codes to strengthen the security and robustness further.
		\item We evaluate the capacity, imperceptibility, and robustness of the proposed method.
		\item We conduct a comparison with related work in an implemented testbed of ten algorithms.
	\end{itemize}
	
	The remaining sections of this paper are structured as follows:
	The necessary background information and nomenclature are introduced in Section~\ref{sec:background}. An overview of similar algorithms and concepts is presented in Section~\ref{sec:related-work}. Our proposed embedding and extraction technique is detailed in Section~\ref{sec:proposed-method}, including information about our reference implementation. The results of a benchmark evaluation are presented in Section~\ref{sec:evaluation} and discussed in Section~\ref{sec:discussion}. The paper concludes in Section~\ref{sec:conclusion}, with additional information in the Appendix.

	\section{Background}
	\label{sec:background}
	This section introduces the terminology and notation needed to understand this work.
	
	\subsection{Watermarking and Steganography}
	Distinguishing between the different terminologies used in the domains of \emph{cryptography} and \emph{information hiding} is crucial and has been discussed comprehensively in the literature~\cite{Ahvanooey.2018,Ahvanooey.2022,Petitcolas.1999,Rizzo.2016,Rizzo.2019}. Starting with the broadest concept, ``information hiding is the science of concealing a secret message or watermark inside a cover media (a host file/message) for providing various security purposes such as content authentication, integrity verification, covert communication, and so on''~\cite[p.~56]{Ahvanooey.2022}. Information hiding can further be divided into steganography and watermarking. It is distinct from cryptography because, at its root, it is not about transforming plain text into an encrypted cipher text~\cite{Ahvanooey.2018}. A detailed survey and classification of information-hiding techniques has been published in~\cite{Petitcolas.1999}.
	
	The process of hiding a secret message inside a cover medium ``in a way that one cannot detect it''~\cite[p.~6367]{Alkawaz.2016} is termed \emph{steganography}. Techniques used to perform steganography can be classified into character-level, bit-level, and hybrid methods~\cite{Krishnan.2017}. Another definition is as follows:  ``Steganography embeds a secret message inside an innocent looking cover medium, stealthily, without creating any attention. The cover medium used can be a text, image, audio, video, network packets, etc.''~\cite[p.~1]{Krishnan.2017}.
	
	In contrast, \emph{digital watermarking} focuses on inserting ``a visible or an invisible, preferably the latter, identification code that permanently is embedded in the data''~\cite[p.~230]{Jalil.2009}. Watermarked data assets can range from images, audio, and video to the under-researched and most difficult cover medium of plain text~\cite{Bertini.2019}, often referred to as digital watermarking, text watermarking, or digital text watermarking. An overview of the definitions that have arisen over time is presented in Table~\ref{tab:definitions}.
	
	\begin{table}[htb]
		\centering
		\caption{Digital watermarking definitions.}\label{tab:definitions}
		\setlength{\tabcolsep}{3pt}
		\begin{tabular}{p{185pt}p{52pt}}
			\hline
			Definition & Reference \\
			\hline
			``A digital watermark can be described as a visible or an invisible, preferably the latter, identification code that permanently is embedded in the data.'' & \cite[p.~230]{Jalil.2009}\\
			``In digital watermarking, relevant information is embedded in an imperceptive way into a digital document. The embedded information is called a watermark.'' & \cite[p.~19]{Qadir.2006}\\
			``Digital  watermarking technology is a typical information hiding method, which covers text, image and video.'' & \cite[p.~1311]{Qi.2023}\\
			\hline
		\end{tabular}
	\end{table}
	
	To bring everything together, Rizzo et al.~\cite[p.~97]{Rizzo.2016} summarize the concepts as follows: ``While cryptography algorithms make unreadable the information by applying a kind of permutation or substitution to the original content, the steganography algorithms provide techniques to hide new information into the carrier, that is a readable document. Whereas watermarking algorithms ensure the authentication and the copyright protection by applying a watermark to the digital content.''
	
	\subsubsection{Classification}
	The existing methods can be further classified according to the following attributes based on~\cite{Petitcolas.1999,Rizzo.2019,Qi.2023}:
	\begin{itemize}
		\item \emph{Cover Medium:} The type of data in which the secret message is embedded, such as image, audio, video, or text files~\cite{Petitcolas.1999}.
		\item \emph{Imperceptibility:} The secret message can be embedded visibly or invisibly inside the cover~\cite{Rizzo.2019,Qi.2023}.
		\item \emph{Blindness:} A blind method allows the secret message to be extracted without knowledge of the original cover, while a non-blind method requires the original cover for extraction~\cite{Rizzo.2019,Qi.2023}.
		\item \emph{Robustness:} A robust method can withstand attacks and intentional or unintentional modifications, whereas a fragile method can not~\cite{Petitcolas.1999,Rizzo.2019}.
	\end{itemize}
	Our proposed method is an invisible and blind method that uses text as the cover medium, aiming for high robustness.

	\subsection{Notation and Unicode Whitespace Characters}
	\label{sec:notations}
	In this subsection, we introduce the notation used in this paper to describe our proposed method, which follows~\cite{Ahvanooey.2020}. All iterable elements in the algorithmic descriptions, like lists, are indexed starting at one. Since the information-hiding scheme is based on plain text, let $u$ be a member of the set $\mathcal{U} := \{u : u$ is Unicode character$\}$ of the 154\,998 characters in version 16.0.0 of the Unicode standard~\cite{TheUnicodeConsortium.2025}. Let $s$ be a Unicode whitespace character with positive width, and let $\mathcal{S} := \{s : s$ is space character $ \land\ s \in U\}$ be the set of all 17 such characters in this version of the standard. A character like the zero-width space (U+200B) ``although called a `space'  in its name, does not actually have any width or visible glyph in display (\dots) and is treated as a format control character, rather than as a space character''~\cite[p.~326]{TheUnicodeConsortium.2025} and, therefore, is not an element of $\mathcal{S}$. We define the classical and most commonly used space character U+0020 as $\delta \in \mathcal{S}$. We initially evaluated every $s$ in~\cite{Hellmeier.2025} to define our own subset of whitespace characters as the alphabet $\mathcal{A}_+ := \{a : a \in S \land a \in U \land a$ meets criteria$\}$, where the criteria are \emph{non-noticeability} for humans and \emph{robustness} in different applications and file formats.
	
	\begin{table}[htb]
		\centering
		\caption{Whitespace evaluation based on~\cite{Korpela.2002}.}\label{tab:whitespace-evaluation}
		\setlength{\tabcolsep}{1.8pt}
		\begin{tabular}{p{60pt}lcccccc}
			\hline
			Name &  Code & Visibil. & .txt & .docx & .pdf & Mail & Teams\\
			\hline
			Space & U+0020                      		& \ding{51} & \ding{51} & \ding{51} & \ding{51} & \ding{51} & \ding{51}\\
			No-Break Space & U+00A0             		& \ding{51} & \ding{51} & \ding{55} & \ding{55} & \ding{55} & \ding{55}\\
			Ogham Space Mark & U+1680           		& \ding{55} & \ding{51} & \ding{51} & \ding{51} & \ding{51} & \ding{51}\\
			En Quad & U+2000                    		& \ding{55} & \ding{51} & \ding{51} & (\ding{51}) & \ding{51} & \ding{51}\\
			Em Quad & U+2001                    		& \ding{55} & \ding{51} & \ding{51} & (\ding{51}) & \ding{51} & \ding{51}\\
			En Space & U+2002                   		& \ding{55} & \ding{51} & \ding{51} & \ding{55} & \ding{51} & \ding{51}\\
			Em Space & U+2003                   		& \ding{55} & \ding{51} & \ding{51} & \ding{55} & \ding{51} & \ding{51}\\
			\textbf{Three-per-Em Space} & U+2004        & \ding{51} & \ding{51} & \ding{51} & (\ding{51}) & \ding{51} & \ding{51}\\
			Four-per-Em Space & U+2005          		& \ding{51} & \ding{51} & \ding{55} & \ding{55} & \ding{55} & \ding{51}\\
			Six-per-Em Space & U+2006           		& (\ding{55}) & \ding{51} & \ding{51} & (\ding{51}) & \ding{51} & \ding{51}\\
			Figure Space & U+2007               		& \ding{55} & \ding{51} & \ding{51} & (\ding{51}) & \ding{51} & \ding{51}\\
			\textbf{Punctuation Space} & U+2008         & \ding{51} & \ding{51} & \ding{51} & (\ding{51}) & \ding{51} & \ding{51}\\
			\textbf{Thin Space} & U+2009                & \ding{51} & \ding{51} & \ding{51} & (\ding{51}) & \ding{51} & \ding{51}\\
			Hair Space & U+200A                 		& (\ding{55}) & \ding{51} & \ding{51} & (\ding{51}) & \ding{51} & \ding{51}\\
			\textbf{Narrow No-Break Space} & U+202F 	& \ding{51} & \ding{51} & \ding{51} & (\ding{51}) & \ding{51} & \ding{51}\\
			\textbf{Medium Mathematical Space} 			& U+205F  & \ding{51} & \ding{51} & \ding{51} & (\ding{51}) & \ding{51} & \ding{51}\\
			Ideographic Space & U+3000          		& \ding{55} & \ding{51} & \ding{51} & \ding{55} & \ding{51} & \ding{51}\\ 
			\hline
		\end{tabular}
	\end{table}
	
	For the \emph{non-noticeability} or visibility criteria, we compared the widths of the whitespace characters, taken from~\cite{Korpela.2002}, with that of the standard space $\delta$ ($\approx 1/4 $~em). If abnormalities are present that cause unusual space, we classify the character as having a different visibility from $\delta$, depicted as ``\ding{55}'' in Table~\ref{tab:whitespace-evaluation}. When the difference is not noticeable, ``\ding{51}'' is shown ($\approx 1/3$ to $1/5$~em), whereas ``(\ding{55})''  indicates that the difference could be noticeable to human eyes (e.g., $1/2$ or $1/6$~em).
	
	For the application and file type \emph{robustness} criteria, we tested the different whitespace characters in text (.txt) files, Microsoft Word files (.docx), PDF files created using Word, emails, and Microsoft Teams Chat. All tests were executed on Windows, Linux (Ubuntu), and macOS for operating system independence. These file types and programs were selected because they are considered industry standards for office and collaboration tools in many fields, except phone calls and SMS~\cite{DataReportal.2023}.

	Only the three-per-em space (U+2004), the punctuation space (U+2008), the thin space (U+2009), the narrow no-break space (U+202F), and the medium mathematical space (U+205F) are not noticed by humans and are robust in most of our tested applications and file formats. Therefore, these form the whitespace homoglyph alphabet $\mathcal{A}_+$; their names are bold in Table~\ref{tab:whitespace-evaluation}. Thus, we can see that $\mathcal{A}_+ \subset \mathcal{S} \subset \mathcal{U}$. Our proposed Innamark technique, introduced in Section~\ref{sec:proposed-method}, uses four of the five elements of $\mathcal{A}_+$ to encode the secret message because one element $\phi \in A_+$ is used as a separator character. We denote the alphabet without the separator character as $\mathcal{A}_-$, where $\mathcal{A}_+ = \mathcal{A}_- \cup \{ \phi \} $ and $\phi \notin \mathcal{A}_-$. Depending on the final application, the modular design of our proposed technique allows the usage of other characters in $\mathcal{A}_+$ if a different set is needed due to framework restrictions or the requirements of specific use cases. A summary of the notation is presented in Table~\ref{tab:notations}.
	
	\begin{table}[htb]
		\centering
		\caption{Overview of notation, following~\cite{Ahvanooey.2020}.}
		\label{tab:notations}
		\setlength{\tabcolsep}{3pt}
		\begin{tabular}{lp{155pt}} 
			\hline
			Symbol & Meaning \\
			\hline
	        $CT$ & Cover text\\
			$SM$ & Secret message text\\
			$SM_{bytes}$ & Byte representation of $SM$ as a sequence of digits\\
			$SM_H$ & Hidden whitespace representation of $SM_{bytes}$\\
			$CT_{SM}$ & Cover text containing a hidden secret message\\
			$\mathcal{U}$ & Set of all Unicode characters\\
			$\mathcal{S}$ & Set of all 17 Unicode space characters\\
			$\delta$ & Classical Unicode whitespace (U+0020)\\
			$\phi$ & Separator whitespace character\\
			$\mathcal{A}_+$ & Whitespace alphabet with $\phi$\\
			$\mathcal{A}_-$ & Whitespace alphabet without $\phi$\\
			$\theta$ & Option parameter for encryption, compression, etc.\\
			$Emb(CT, SM_{bytes}, \theta)$ & Embedding algorithm\\
			$Ext(CT_{SM})$ & Extraction algorithm\\
			\hline
		\end{tabular}
	\end{table}
	
	\section{Related Work}
	\label{sec:related-work}
	Over time, a variety of information-hiding algorithms and implementations for watermarking and steganography have been published. Due to the increasing diversity of methods, cover media, and applications, several literature reviews and surveys have been published to organize the cluttered research and application landscape. A more detailed discussion of selected methods is provided in section \ref{sec:related-methods}.
	
	Bender et al.~\cite{Bender.1996} provided one of the first comprehensive overviews of data-hiding methods for image, audio, and text files. Later on, Petitcolas et al.~\cite{Petitcolas.1999} published a survey on information-hiding techniques, focusing on steganography, watermarking, and fingerprinting, including information about possible attacks and a basic overall theoretical framework. More specialized overviews with solutions focusing on text steganography have been provided by Ahvanooey et al.~\cite{Ahvanooey.2019}, Krishnan et al.~\cite{Krishnan.2017}, and Majeed et al.~\cite{Majeed.2021}. Current challenges are discussed by Ahvanooey et al.~\cite{Ahvanooey.2022} and Tyagi et al.~\cite{Tyagi.2023}, with the latter considering concrete application possibilities.
	
	In addition to reviews, researchers have formally compared existing methods to identify their strengths and weaknesses. Ahvanooey et al.~\cite{Ahvanooey.2018b} compared watermarking and steganography methods by differentiating their embedding techniques and evaluating them according to the criteria of imperceptibility, embedding capacity, robustness, security, and computational cost~\cite{Ahvanooey.2018b}. One of the latest evaluations of text steganography methods was published by Knöchel and Karius~\cite{Knochel.2024}, who compared their capacity, imperceptibility, robustness, and complexity with a specialized focus on malware.
	
	\subsection{Related Methods}
	\label{sec:related-methods}
	To situate our proposed Innamark method within the research landscape, we consider the most relevant text watermarking and steganography methods. In the \ac{LLM} problem domain initially set out, Kirchenbauer et al.~\cite{Kirchenbauer.2023} and Christ et al.~\cite{Christ.2024b} presented token-based watermarking schemes, with the latter focusing on undetectability, completeness, and soundness. These ideas were integrated into SynthID, the watermarking engine used by Google's Gemini \ac{LLM}~\cite{Dathathri.2024}. Steinebach~\cite{Steinebach.2024} generated a text based on sets of letters. These methods are classified as linguistic since they make use of \ac{LLM} text generation. Such methods are problematic for cover texts whose semantics are essential. Thus, we focus on format-based methods that use insertion- or substitution-based embedding techniques~\cite{Knochel.2024}. Other types of format-based methods, as well as linguistic and random or statistical generation methods~\cite{Majeed.2021,Knochel.2024}, are not considered further since either they do not work on plain text documents due to the lack of formatting options or they change the semantics or structure of the cover text. We implemented all the methods presented here for our benchmark evaluation in Section~\ref{sec:evaluation}. A summary is provided in Table~\ref{tab:related-methods}.
	
	\subsubsection{SNOW}
	One of the oldest whitespace steganography methods for ASCII texts is \ac{SNOW}~\cite{Kwan.2013}. Although the first release goes back to the 20th century, the last update, with a change to the open-source Apache 2.0 license, was made in 2013. It has Java and Windows DOS versions and a C implementation last updated in 2016~\cite{Kwan.2016}. The embedding process encodes the secret message into tab and space characters and appends it to the cover text, starting with a tab character under the consideration of a predefined line length~\cite{Kwan.2013}. Upstream compression and encryption can be enabled before the encoding process.
	
	\subsubsection{UniSpaCh}
	A well-known algorithm in the field of information hiding for text documents is UniSpaCh, proposed by Por et al.~\cite{Por.2012}. It is an extended version of WhiteSteg, which replaces a single whitespace character between two words or paragraphs with either one or two characters to encode a zero or one~\cite{Por.2008}. UniSpaCh uses two different methods to embed the secret message in the text. For spaces between words and sentences, regular whitespace characters either remain as they are or are extended by adding a thin, six-per-em, or hair space to encode two bits per embedding location~\cite{Por.2012}. For end-of-line and inter-paragraph spacings, the remaining space is filled with a combination of hair, six-per-em, punctuation, and thin spaces to encode two bits per character~\cite{Por.2012}.
	
	\subsubsection{AITSteg}
	Ahvanooey et al.~\cite{Ahvanooey.2018} proposed a text steganography technique for SMS or social media communication. The embedding method transforms the secret message into zero-width characters with a Gödel function and uses the sending/receiving time and the length of the secret message to insert it before the cover text.
	
	\subsubsection{Shiu et al.}
	The data hiding method proposed by Shiu et al.~\cite{Shiu.2018} focuses on communication over messengers of social media networks. Due to the small width of social media messaging windows, the method is based on a fixed line length and can hide three bits per line of the cover text~\cite{Shiu.2018}. After encoding a secret message into a bit stream based on the ASCII mapping, it embeds the first bit by adding a whitespace at the end of a line, changing the length of the line to embed the second bit, and adding a whitespace between two words to embed the third bit~\cite{Shiu.2018}.
	
	\subsubsection{Rizzo et al.}
	A text-watermarking technique based on replacing Unicode characters with specific confusables, also known as homoglyphs, was initially proposed in~\cite{Rizzo.2016} and extended to a fine-grain watermarking approach in~\cite{Rizzo.2019}. The latter method generates a watermark by using a keyed hash function with a secret message as a watermark and a secret password~\cite{Rizzo.2019}. Afterward, the watermark is embedded in the cover text by replacing specific characters with their confusables or leaving them unchanged to embed one bit in each and replacing spaces with a set of specific whitespace characters to embed three bits in each~\cite{Rizzo.2019}.
	
	\subsubsection{StegCloak}
	The open-source implementation StegCloak published by~\cite{KuroLabs.2020}, as described in~\cite{Mohanasundar.2020}, is a JavaScript steganography tool that is able to hide a secret message inside a cover text with optional password encryption and \ac{HMAC}. In the embedding process, the secret message is compressed, optionally encrypted, and encoded in a set of zero-width characters to be inserted in one location after a classical whitespace of the cover text~\cite{Mohanasundar.2020}.
	
	\subsubsection{Lookalikes}
	Another implementation is the Unicode Lookalikes algorithm by~\cite{Thompson.2021} as part of the Python package pyUnicodeSteganography. Similar to~\cite{Rizzo.2019}, the method replaces specific characters with their confusables to encode a secret message inside the cover text~\cite{Thompson.2021}.
	
	\subsubsection{CovertSYS}
	Ahvanooey et al.~\cite{Ahvanooey.2022b} presented a multilingual steganography method focusing on short messages in social media networks. Like the previous approach in~\cite{Ahvanooey.2018}, four zero-width characters and a timestamp are used to encode the secret message. Further, a password-based approach with a \ac{OTP}  and an XOR operation are used to transform the secret message into an encrypted bit stream that is appended to the cover text~\cite{Ahvanooey.2022b}.
	
	\subsubsection{Shazzad-Ur-Rahmen et al.}
	The data-hiding approach of Shazzad-Ur-Rahmen et al.~\cite{ShazzadUrRahman.2021} can embed five bits per embeddable location, whereas their updated version~\cite{ShazzadUrRahman.2023} can embed six. The main idea of the latter procedure is to encrypt the secret message using AES and convert the resulting binary stream into blocks of six bits~\cite{ShazzadUrRahman.2023}. With the help of two lists, specific Unicode characters are replaced with their confusables, and whitespace characters are replaced with a particular combination of smaller whitespace characters to embed the secret message in the cover text~\cite{ShazzadUrRahman.2023}.
	
	\begin{table*}[!htb]
		\centering
		\caption{Overview of Related Methods.}
		\label{tab:related-methods}
		\setlength{\tabcolsep}{4pt}
		\begin{tabular}{p{136pt}p{42pt}p{80pt}p{220pt}}
			\hline
			Name & Release & Publication & Techniques \\
			\hline
        	SNOW~\cite{Kwan.2013,Kwan.2016} & Before 1998 & Docum., Software & Appending additional tabs and whitespace characters at the end of the text.\\
			UniSpaCh~\cite{Por.2012} & 2012 & Paper & Adds small whitespace characters between words and sentences and fills up lines and inter-paragraph spacings.\\
			AITSteg~\cite{Ahvanooey.2018} & 2018 & Paper & Hides data at the beginning of the cover text by using symmetric-key encoding and a transformation into zero-width characters.\\
			Shiu et al.~\cite{Shiu.2018} & 2018 & Paper & Hides data line-wise by either changing the line length or adding whitespace between words or at the end of the line.\\
			Rizzo et al.~\cite{Rizzo.2016,Rizzo.2019} & 2016/2019 & Paper & Replaces whitespace and other Unicode characters with their confusables.\\
			Steg\-Cloak~\cite{KuroLabs.2020,Mohanasundar.2020} & 2020 & Blog post, Software & Inserts the secret message at one location in the cover text using zero-width characters.\\
			Looka\-likes~\cite{Thompson.2021} & 2021 & Software & Replaces a specific set of Latin characters with their Unicode confusables.\\
			Covert\-SYS~\cite{Ahvanooey.2022b} & 2022 & Paper & Adds zero-width characters at the end of the cover text by using the current date and time and an \ac{OTP}.\\
			Shazzad-Ur-Rahman et al.~\cite{ShazzadUrRahman.2021,ShazzadUrRahman.2023} & 2021/2023 & Paper & Replaces confusables and changes whitespace to a specific combination of small whitespace characters.\\
			Innamark~\cite{Hellmeier.2025,FraunhoferISST.2025} & 2025 & Paper, Software & Replaces whitespaces with a specific set of robust, similar-looking whitespace characters.\\
			\hline
		\end{tabular}
	\end{table*}

	\section{Proposed Method}
	\label{sec:proposed-method}
	In this section, we present Innamark, our invisible and blind information-hiding technique for plain text. Since existing methods either lack robustness in some applications, increase the number of characters, or are recognizable by humans, they are unsuitable for the \ac{LLM} use case described in Section~\ref{sec:introduction}. Therefore, this section presents the embedding and extraction method in Algorithm~\ref{alg:emb} and Algorithm~\ref{alg:ext} based on our nomenclature introduced in Section~\ref{sec:notations}. It further includes a concrete example based on a \emph{Lorem ipsum} dummy text and information about our implemented prototypes.
	
	\subsection{Embedding}
	\label{sec:embedding}
	The proposed embedding method, detailed in Algorithm~\ref{alg:emb}, can hide any byte-encoded sequence in a Unicode-encoded cover text $CT$. The examples in this paper are based on a secret message $SM$ in text form, in which every character is transformed into its UTF-8 byte representation $SM_{bytes}$.
	
	\begin{algorithm}
		\caption{Embedding $Emb(CT, SM_{bytes}, \theta)$}
		\label{alg:emb}
		\begin{algorithmic}[1]
			\renewcommand{\algorithmicrequire}{\textbf{Input:}}
			\renewcommand{\algorithmicensure}{\textbf{Output:}}
			\REQUIRE Cover text ($CT := \{c_1, c_2, \dots, c_n\}$, $\forall c \in \mathcal{U}$);\\
			Secret message ($SM_{bytes} := \{q_1, q_2, \dots, q_n\}$, $\forall q \in \{0, 1, \dots, 255\}$);\\
			Configuration parameter ($\theta$)
			\ENSURE Cover text with hidden secret message ($CT_{SM}$)
			
			\STATE \textit{$\triangleright$ Insert tag}
			\STATE $SM_{bytes}, SM_H \gets applyTag(SM_{bytes}, \theta)$
			
			\STATE \textit{$\triangleright$ Encode secret message}
			\STATE $d \gets \left\lceil\frac{\log_2 2^8}{\log_2 |\mathcal{A}_-|}\right\rceil$
			\FOR {\textbf{each} $q \in SM_{bytes}$}
			\FOR {$i \gets 1$ \TO $d$}
			\STATE $r \gets q \mod |\mathcal{A}_-|$
			\STATE $q \gets \bigl\lfloor\frac{q}{|\mathcal{A}_-|}\bigr\rfloor$
			\STATE $SM_H \gets SM_H + a_{r+1}$ \hfill \textit{$\triangleright$ $a_{r+1} \in \mathcal{A}_-$}
			\ENDFOR
			\ENDFOR
			
			\STATE \textit{$\triangleright$ Insert secret message}
			\STATE $i \gets 0$
			\FOR {\textbf{each} $c \in CT$}
			\IF {$c = \delta$}
			\IF {$i = 0$}
			\STATE $CT_{SM} \gets CT_{SM} + \phi$
			\STATE $i \gets i + 1$
			\ELSIF {$0 < i \le |SM_H|$}
			\STATE $w_{H_i} \gets SM_{H_i}$ \hfill \textit{$\triangleright$ $w_{H_i} \in \mathcal{A}_-$}
			\STATE $CT_{SM} \gets CT_{SM} + w_{H_i}$
			\STATE $i \gets i + 1$
			\ELSE
			\STATE $i \gets 0$
			\ENDIF
			\ELSE
			\STATE $CT_{SM} \gets CT_{SM} + c$
			\ENDIF
			\ENDFOR
			\RETURN $CT_{SM}$
		\end{algorithmic}
	\end{algorithm}
	
\begin{figure}[!htb]
		\centering
		\includegraphics[width=1.0\columnwidth]{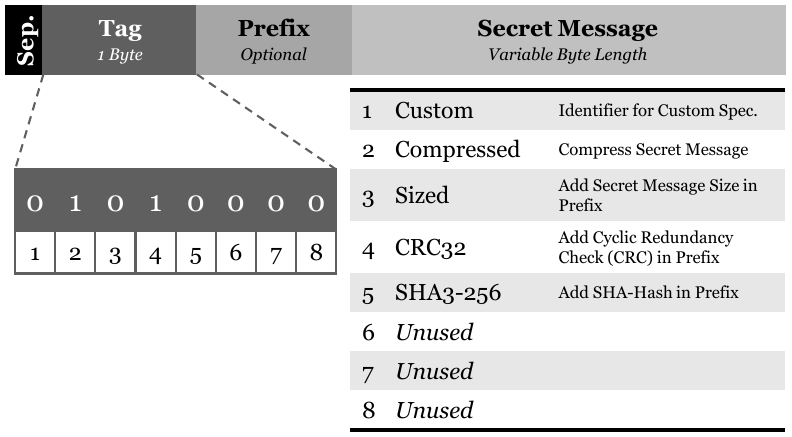}
		\caption{InnamarkTag structure.} \label{fig:innamarktag-structure}
	\end{figure}

	The embedding function $Emb(CT, SM_{bytes}, \theta)$ begins by analyzing the configuration parameter $\theta$. This specifies the type of the secret message, which we call a \emph{InnamarkTag}. It offers optional functionalities that can be enabled in any combination by the end user, namely, encryption, compression, hashing, and error-correcting codes. Depending on the user's choice, $\theta$ defines the type used to calculate the \emph{tag}. The tag has a fixed length of one byte and is returned by the $applyTag$ method with the updated secret message, depending on the user's choice. Each bit in the tag indicates whether an option, such as compression, is enabled ($1$) or not ($0$). Thus, the tag describes the format of the secret message, similar to tags or headers in network packets~\cite{Deering.2017}. Fig.~\ref{fig:innamarktag-structure} provides an overview of the InnamarkTag structure, the definition of each bit in the tag, and an example of a tag with enabled compression (second compression bit set to one) and error correction (fourth CRC32 bit set to one).
	
	An InnamarkTag starts after the separator character $\phi$, with the tag having a fixed length of one byte, followed by an optional prefix whose content depends on the tag, followed by the secret message itself.
	
	Next, the encoding process starts transforming the secret message with the InnamarkTag and optional prefixes $SM_{bytes}$ into a sequence of whitespace characters $SM_H$ from our homoglyph alphabet $\mathcal{A}_-$. Since $|\mathcal{A}_-| = 4$, where $|\cdot|$ denotes the cardinality, each byte of the secret message is represented by four elements of $\mathcal{A}_-$ because
	\begin{equation}
		\left\lceil\frac{\log_2 2^8}{\log_2 |\mathcal{A}_-|} \right\rceil = 4. 
	\end{equation}
	To fully include $SM_H$ in a cover text $CT$, the number of standard space characters $\delta$ in $CT$ must be at least the number of elements of the hidden secret message $SM_H$:
	\begin{equation}
		|\{x \in CT : x = \delta\}| \geq |SM_H|.
	\end{equation}
	
	The final version of the cover text with the hidden secret message $CT_{SM}$ is created by replacing all $\delta$ successively with the elements of $SM_H$. If, on the one hand, the input cover text $CT$ does not have any normal whitespace $\delta$, e.g., if the algorithm has already been applied to it, it is not possible to embed the secret message. If, on the other hand, $CT$ has more $\delta$ characters than $|SM_H|$, the insertion process starts again until all $\delta$ are replaced to include the secret message multiple times. Thus, the resulting text does not contain standard space characters:
	\begin{equation}
		\forall x \in CT_{SM} : x \in \mathcal{U} \land x \neq \delta.
	\end{equation}
	The use of multiple insertions improves the robustness of the information-hiding scheme, as changes do not necessarily destroy the secret message. The entire embedding algorithm $Emb(CT, SM_{bytes}, \theta) = CT_{SM}$ is presented in Algorithm~\ref{alg:emb}.
	
	\begin{figure}[htb]
		\centering
		\includegraphics[width=1.0\columnwidth]{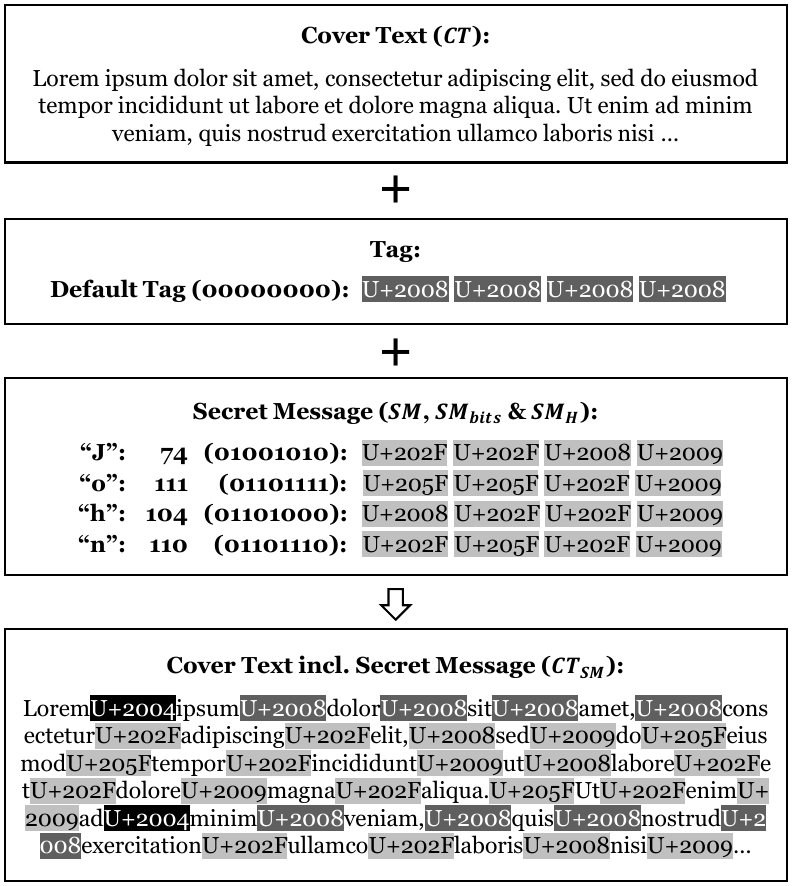}
		\caption{Example embedding with a default tag.}
		\label{fig:embedding-example}
	\end{figure}
	
\begin{figure}[htb]
		\centering
		\includegraphics[width=0.8\columnwidth]{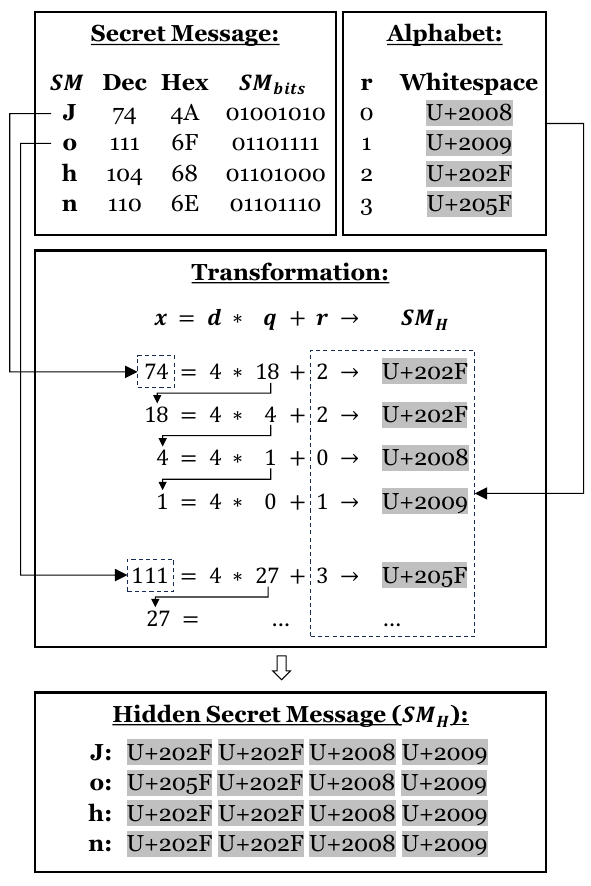}
		\caption{Example of transforming a secret message into the whitespace alphabet.}
		\label{fig:transformation-example}
	\end{figure}

	An example of the proposed embedding algorithm is shown in Fig.~\ref{fig:embedding-example} for the process and in Fig.~\ref{fig:innamark-display-example} for input/output comparison. The secret message $SM = $ ``John'' is to be hidden inside the cover text $CT =$ ``Lorem ipsum \dots'' with an empty configuration parameter $\theta$ to use the default InnamarkTag without compression, hashing, or error correction. Each character of $SM$ is encoded into its byte representation $SM_{bytes}$ and transformed into the whitespace alphabet. In this case, the first letter ``J'' of the secret message is represented by the UTF-8 hexadecimal value 4A (U+004A), which equals the decimal value 74. Afterward, each value of $SM_{bytes}$ is encoded into $SM_H$ by transforming it into the alphabet $\mathcal{A}_-$ with a loop-based modulo operation, as described in Algorithm~\ref{alg:emb}. This uses the length of the alphabet without separator characters as the divisor $d = |\mathcal{A}_-| = 4$, as illustrated in Fig.~\ref{fig:transformation-example}. The remainder of each division operation indicates the index of the whitespace in $\mathcal{A}_-$ needed to build the complete sequence of whitespace characters $SM_H$ as a representation of the secret message $SM$. Next, each whitespace character $\delta$ of $CT$ is replaced with the corresponding space in $SM_H$. Since $CT$ has more whitespace characters than needed, the insertion process starts again with the separator character $\phi$, represented by the black U+2004 in Fig.~\ref{fig:embedding-example}. Afterward, it inserts the default tag and the first character ``J'' of the secret message a second time and stops after all whitespace characters have been replaced.
	
	\subsection{Extraction}
	\label{sec:extraction}
	The overall extraction method $Ext(CT_{SM}) = SM_{bytes}$ is split into three parts: extraction of the encoded message, tag analysis, and decoding. The first part starts by iterating over all characters of the input text, including the secret message $CT_{SM}$, until it finds the first occurrence of $\phi$ as the separator character. Through the filtering of $\mathcal{A}_+$, it extracts the full InnamarkTag, an optional prefix, and a hidden secret message $SM_H$, as specified in Fig.~\ref{fig:innamarktag-structure}.
	
	The second part analyzes and evaluates the entire InnamarkTag by calling $analyzeTag()$. If hashing is enabled, it checks and verifies the hash of $SM_H$ and returns an error if problems occur. It can also decompress the message, check the size, or apply the CRC32 prefix.
	
	The third part decodes the hidden secret message $SM_H$ into its byte representation $SM_{bytes}$. The step size for the decoding part depends on the length of the alphabet without the separator character and is defined as
	\begin{equation}
		d := \left\lceil \frac{\log_2 2^8}{\log_2 |\mathcal{A}_-|} \right\rceil
	\end{equation}
	with $d = 4$ for our alphabet because each byte is represented by four whitespace characters from $\mathcal{A}_-$. The result of the cascading modulo operation from Algorithm~\ref{alg:emb} can be transformed back into its byte representation $b$. All $b$ form the secret message representation $SM_{bytes}$, which in turn can be converted into the UTF-8 representation of the decoded secret message text $SM$. The overall extraction process is summarized in Algorithm~\ref{alg:ext}.
	
	\begin{algorithm}
		\caption{Extraction $Ext(CT_{SM})$}
		\label{alg:ext}
		\begin{algorithmic}[1]
			\renewcommand{\algorithmicrequire}{\textbf{Input:}}
			\renewcommand{\algorithmicensure}{\textbf{Output:}}
			\REQUIRE Cover text with a hidden secret message ($CT_{SM}$)
			\ENSURE Extracted secret message bytes ($SM_{bytes}$) or error
			
			\STATE \textit{$\triangleright$ Extract secret message}
			\FOR {\textbf{each} $c \in CT_{SM}$}
			\IF {$c \in \mathcal{A}_+$}
			\IF {$c = \phi$ \textbf{and} $SM_H \neq \emptyset$}
			\STATE \textbf{break}
			\ELSIF{$c \in \mathcal{A}_-$}
			\STATE $SM_H \gets SM_H + c$
			\ENDIF
			\ENDIF
			\ENDFOR
			
			\STATE \textit{$\triangleright$ Analyze tag and prefix}
			\STATE $SM_H, error \gets analyzeTag(SM_H)$
			\IF {error}
			\RETURN error
			\ENDIF
			
			\STATE \textit{$\triangleright$ Decode secret message}
			\STATE $d \gets \left\lceil\frac{\log_2 2^8}{\log_2 |\mathcal{A}_-|}\right\rceil$
			\FOR {$i \gets 0$ \TO $|SM_{H}|$ \text{step} $d$}
			\FOR {$y \gets 0$ \TO $d-1$}
			\STATE $a_{r+1} \gets SM_{H_{i+y+1}}$ \hfill \textit{$\triangleright$ $a_{r+1} \in \mathcal{A}_{-}$}
			\STATE $b \gets b + r \cdot d^y$ \hfill \textit{$\triangleright$ $r \in [0,\dots,d-1]$}
			\ENDFOR
			\STATE $SM_{bytes} \gets SM_{bytes} + b$
			\ENDFOR
			\RETURN $SM_{bytes}$
		\end{algorithmic}
	\end{algorithm}
	
	\subsection{Implementation}
	The watermark embedding and extraction methods were implemented as a generic library in the Kotlin programming language to test and validate them. Kotlin was chosen since it is interoperable with the widely used Java programming language while supporting multiplatform targets. Thus, our implementation can be used in applications supporting the \ac{JVM} and in JavaScript solutions due to the availability of both build targets.
	
	To test the library, we developed a \ac{CLI} tool for the \ac{JVM} that can embed and extract a byte-encoded string into another string or text-based document, such as a plain text file. Additionally, a web interface was implemented as a second usage example for the JavaScript build target. This front end acts as a graphical user interface and is likewise able to embed and extract watermarks in cover texts.
	
	The source code of our implementations is made publicly available in~\cite{FraunhoferISST.2025} to ensure full transparency and applicability.
	
	\section{Evaluation and Experimental Results}
	\label{sec:evaluation}
	To analyze and evaluate our proposed Innamark technique, this section compares it with state-of-the-art methods for text watermarking and steganography.
	
	Several empirical research methodologies are employed in software engineering, including simulations, benchmarks, case studies, and controlled experiments~\cite{Hasselbring.2021}. We use benchmarking in this evaluation since it is a ``standard tool for the competitive evaluation and comparison of competing systems or components according to specific characteristics''~\cite[p.~333]{Kistowski.2015}. Numerous types of benchmarks exist, but we focus on \emph{specification-based benchmarks} since they concentrate on a business problem and require development work before running the benchmark~\cite{Kistowski.2015,Kounev.2020}.
	
	We based our benchmark on criteria from existing comparisons and evaluations because benchmarks should be developed by the community instead of a single researcher~\cite{Hasselbring.2021, Sim.2003}. This leads to the following set of criteria, also used in related work~\cite{Majeed.2021,Ahvanooey.2018b,Knochel.2024}:
	\begin{itemize}
		\item \textbf{Capacity:} Describes the embedding amount as a relationship between the length of the secret message and the cover text.
		\item \textbf{Imperceptibility:} Also known as invisibility, refers to the visual and perceived differences between a text with a secret message and one without.
		\item \textbf{Robustness:} A broad term mainly focuses on how reliably the secret message stays inside the cover text in different environments or when attacks are carried out.
	\end{itemize}
	Runtime or execution speed is not benchmarked since the evaluation aims to differentiate the methods according to their core properties rather than efficiency. In the following, we introduce our experimental setup and dataset and present the results of an evaluation conducted for each criterion.
	
	\subsection{Experimental Setup and Dataset}
\begin{figure}[htb]
		\centering
		\includegraphics[width=1.0\columnwidth]{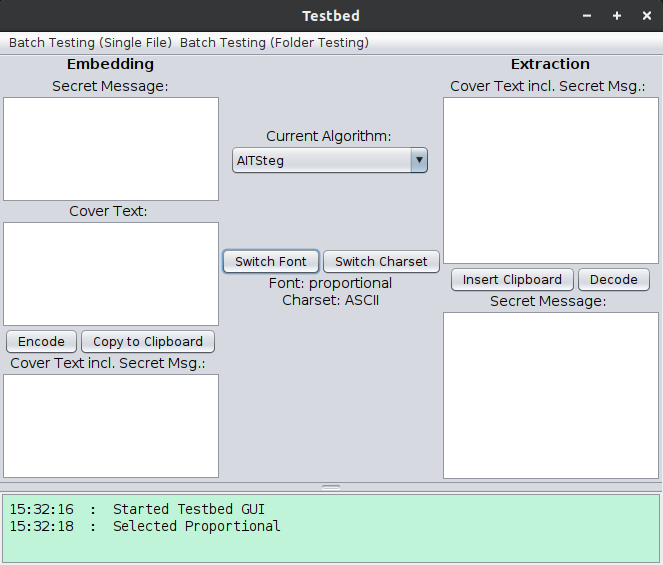}
		\caption{Evaluation \ac{GUI} for embedding and extraction.}
		\label{fig:evaluation-gui}
	\end{figure}

	We created a testbed and implemented all relevant existing methods presented in Section~\ref{sec:related-methods} and Table~\ref{tab:related-methods} in the Java programming language on the basis of the published descriptions, reference implementations, and examples. To ensure a uniform basis and comparability without unnecessary overhead, we consider the embedding and extraction methods only. Optional functionalities like encryption or compression are excluded since they can be applied upstream to all algorithms. Fig.~\ref{fig:evaluation-gui} shows our implemented evaluation \ac{GUI}, consisting of the embedding functionality on the left side, a drop-down box to select the method in the middle with some additional options, and the extraction tool on the right. 
	
	For the benchmark evaluation, we used a large dataset of 1\,000\,000 random English Wikipedia articles as cover text. We tried to embed the largest possible secret message in each of the articles. For imperceptibility and robustness, we applied each algorithm in two execution runs on the 1\,000\,000 cover texts. The first run tried to hide a short four-character secret message inside the dataset, whereas the second run tried to hide a long 455-character secret message. The different lengths and forms of the Wikipedia cover texts, as well as the two different lengths of secret messages, ensure a comprehensive evaluation. More details about the data and evaluation process for transparency and reproducibility are provided in the Appendix.
	
	\subsection{Capacity}
	\label{sec:capacity}
	The embedding capacity analysis provides information about the relationship between the length of the cover text and that of the longest secret message that can be embedded within it. We distinguish the algorithms in Table~\ref{tab:related-methods} into two types:
	\begin{enumerate}
		\item \textbf{Bounded-capacity algorithms:} These have a limited embedding capacity depending on the length, structure, and used characters in the cover text and secret message, primarily due to specific replacements or insertions.\\
		\textbf{Algorithms:} Shiu et al., Innamark, Rizzo et al., Lookalikes, Shazzad-Ur-Rahman et al., UniSpaCh.
		\item \textbf{Unbounded-capacity algorithms:} These have no general limit on their embedding capacity, primarily due to the use of (zero-width) characters. Embedding capacity restrictions only apply if the text length is specified, like SMS or X (formerly Twitter) posts.\\
		\textbf{Algorithms:} AITSteg, CovertSYS, StegCloak, SNOW.
	\end{enumerate}
	
	In our testbed, we analyzed all bounded-capacity algorithms by executing each on the dataset of 1\,000\,000 Wikipedia articles. This process started by hiding a one-byte secret message inside the cover and then repeatedly increasing the length of the secret message until an error occurred, yielding the maximum number of embeddable bytes. This process cannot be executed for the unbounded-capacity algorithms, which can theoretically embed a message of any length inside a given cover text.
	
\begin{figure}[htb]
		\centering
		\includegraphics[width=1.0\columnwidth]{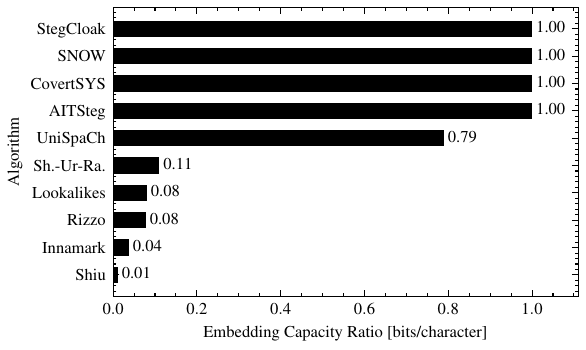}
		\caption{Maximum capacity evaluation results (higher values are better).} \label{fig:capacity}
	\end{figure}

	We based the calculations on the approach described by Rizzo et al.~\cite[p.~13]{Rizzo.2019}: ``The embedding capacity is computed as the average ratio between the number of embedded bits and the number of characters in each document.'' Our dataset has an average cover text size of \around2514 characters. The resulting capacity ratios are depicted in Fig.~\ref{fig:capacity}.
	
	We assign all unbounded-capacity algorithms a value of $1.0$ to illustrate the unprescribed limit. Comparing the others, we find that Shiu et al.\ has the lowest embedding capacity, with $26/2514 \approx 0.01$ bits/character, and that UniSpaCh has the highest of $1983/2514 \approx 0.79$ bits/character. The other four bounded-capacity algorithms are close to each other and have an embedding capacity of around and below 0.1 bits/character, whereas our proposed Innamark method has a capacity of $93/2514 \approx 0.04$ bits/character.
	
	In a direct comparison, our results partly differ from previous studies because the benchmark evaluation depends strongly on the structure, format, and length of the input cover text and the secret message. In~\cite{Rizzo.2019}, a similar analysis with a different cover text dataset based on the New York Times Corpus resulted in an embedding capacity of 0.321 bits/character for UniSpaCh. UniSpaCh embeds the secret message between words, sentences, and paragraphs, with a significant amount hidden between paragraphs. Since~\cite{Rizzo.2019} only used cover texts with a single paragraph as input, UniSpaCh's strength in embedding between paragraphs was not considered in that work.
	
	\subsection{Imperceptibility}
	\label{sec:imperceptibility}
\begin{figure}[htb]
		\centering
		\includegraphics[width=1.0\columnwidth]{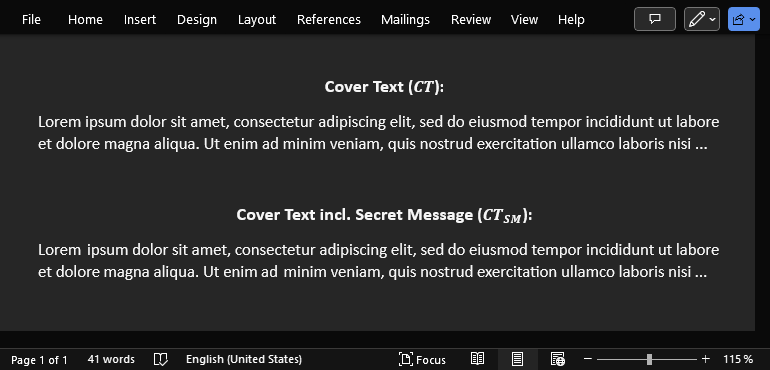}
		\caption{Innamark comparison example in Microsoft Word.} \label{fig:innamark-display-example}
	\end{figure}

	Imperceptibility or invisibility is the ability for a secret message to be concealed in a cover document without causing any visible abnormalities~\cite{Ahvanooey.2018}. Therefore, this benchmark is more relevant for steganography use cases than watermarking~\cite{Por.2012, Rizzo.2016}. It is essential to distinguish between imperceptibility to humans and to machines, which can detect differences, for example, by using statistical metrics. Due to the varying perceptions of humans, this benchmark ``is the most subjective of all the metrics''~\cite[p.~121]{Knochel.2024}. Fig.~\ref{fig:innamark-display-example} shows an illustrative example of our Innamark algorithm as a direct comparison of a plain cover text and a cover text with the embedded secret message ``John'' in Microsoft Office Word version 2408 using the default font ``Calibri (Body)'' in font size 11. To further analyze and compare the imperceptibility of all algorithms, we use four different measurement metrics in the following, namely:
	\begin{enumerate}
		\item the Jaro--Winkler Similarity;
		\item the number of characters;
		\item the file size;
		\item caret navigation.
	\end{enumerate}
	
	\subsubsection{Jaro--Winkler Similarity}
	A standard numerical measurement used to compare the similarity between two character sequences is the \emph{Jaro--Winkler similarity}, also known by the misleading name ``Jaro--Winkler distance''~\cite{Keil.2019}. It is often used to evaluate information-hiding techniques (see~\cite{Ahvanooey.2022b,Ahvanooey.2019,Ahvanooey.2020,Majeed.2022,Majeed.2024,Baawi.2019}). The benchmark is based on the Jaro string comparator $\Phi$, for which a value of $1$ indicates that two strings $s_1$ and $s_2$ are identical and $0$ indicates that the strings have no common characters~\cite{Winkler.1990} :
	\begin{equation}\label{eq:jaro}
		\Phi(s_1,s_2) = 
		\begin{cases} 
			1 & : s_1 = s_2 \\
			\frac{1}{3} \left( \frac{c}{|s_1|} + \frac{c}{|s_2|} + \frac{c - \tau}{c} \right) & \text{: } m >0 \\
			0 & : \text{otherwise}.
		\end{cases}
	\end{equation}
	Here, $|s_1|$ and $|s_2|$ are the lengths of the strings, $c$ is the number of matching characters, $\tau$ is the number of transpositions based on the characters and $m$ for all matching characters~\cite{Keil.2019,Winkler.1990}. The newer Jaro--Winkler similarity $\Phi_n$ from~\cite{Winkler.1990} builds on the Jaro similarity:
	\begin{equation}\label{eq:jaro-winkler}
		\Phi_n(s_1,s_2) = \Phi(s_1,s_2) + i \cdot 0.1 \cdot (1 - \Phi(s_1,s_2)).
	\end{equation}
	It adds the scaling factor $0.1$ and a prefix length $i$ to compare the first characters of the strings~\cite{Winkler.1990}.
	
\begin{figure}[htb]
		\centering
		\includegraphics[width=1.0\columnwidth]{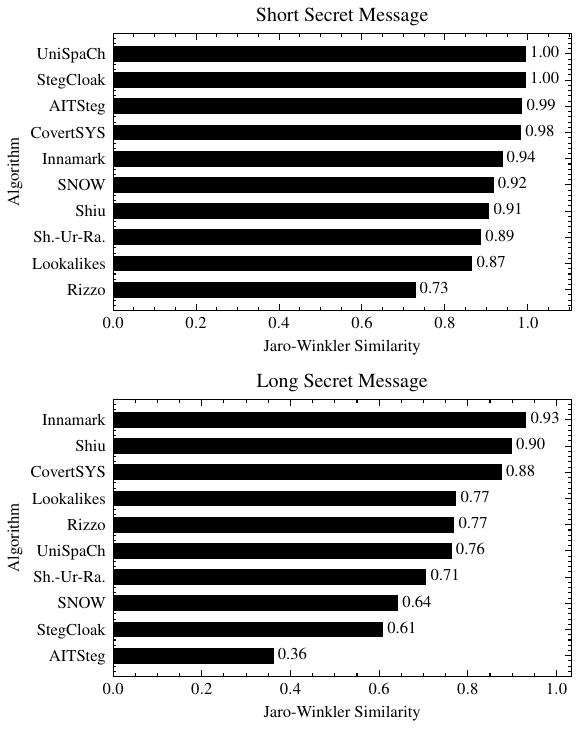}
		\caption{Jaro--Winkler similarity evaluation results (higher values are better).} \label{fig:jaro-winkler}
	\end{figure}

	In our testbed, we used a Java implementation in version 1.12.0 of the Apache Commons Text package~\cite{TheApacheSoftwareFoundation.2025} to calculate the Jaro--Winkler similarity of a plain cover text and a text with an integrated secret message. Like in~\cite{Knochel.2024}, Fig.~\ref{fig:jaro-winkler} shows the average Jaro--Winkler similarity for each algorithm and for both short (four-character) and long (455-character) secret message execution runs. For short secret messages, UniSpaCh shows the best results with a Jaro--Winkler similarity of $\Phi_n = 0.996$, whereas Rizzo et al.\ has the worst result with $\Phi_n = 0.729$. Our proposed Innamark technique has the best similarity of $\Phi_n = 0.931$ for longer messages, whereas AITSteg has the lowest similarity value of $\Phi_n = 0.362$. It is noteworthy that the unbounded algorithms without capacity restrictions have a comparatively high difference in $\Phi_n$ between short and long secret messages.
	
	\subsubsection{Number of Characters}
	A change in the number of characters can reveal that a document has embedded hidden content, leading to poorer imperceptibility. This may be detected automatically by software or by a human, for example, when submitting a text document to a publisher that imposes a character limit.
	
\begin{figure}[htb]
		\centering
		\includegraphics[width=1.0\columnwidth]{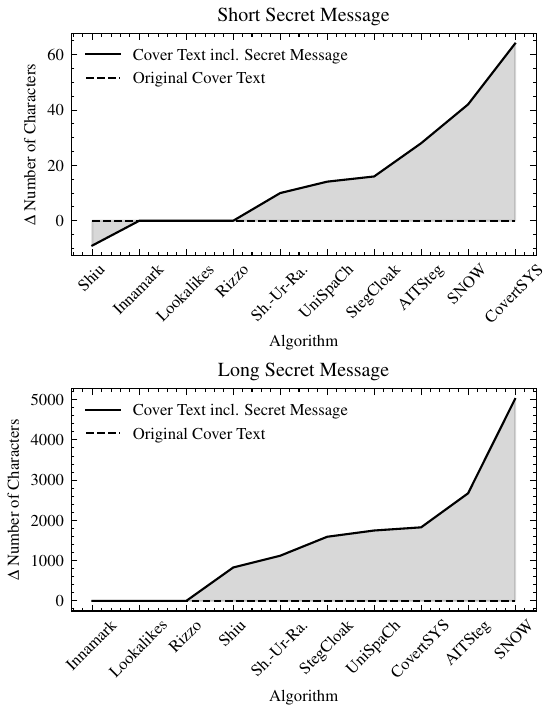}
		\caption{Character size evaluation results (smaller absolute values are better).} \label{fig:character-size}
	\end{figure}

	In our testbed, we compared the number of characters of the original cover text with a text that included an embedded secret message. The mean absolute differences $\Delta$ computed across the dataset are depicted for both long and short messages in Fig.~\ref{fig:character-size}.
	
	Only the pure one-to-one replacement techniques Innamark, Lookalikes, and Rizzo et al.\ do not show a difference. The four unbounded algorithms and UniSpaCh show a significant increase in the number of characters due to the addition of zero-width or small whitespace characters, making them recognizable. The negative value of $\Delta = -8.957$ for Shiu et al.\ is caused by the design of the hiding algorithm. It replaces whitespace characters after a specific length with newlines and removes formatting characters like tabulators, decreasing the number of characters for short secret messages.
	
	\subsubsection{File Size}
	Humans and systems can detect a text file with a hidden secret message by its increased file size. In particular, suspicions may be raised if a file with a small amount of text has a large file size. Thus, Majeed et al.~\cite{Majeed.2021} argue for developing data-hiding methods that create results with minimal file sizes.
	
\begin{figure}[htb]
		\centering
		\includegraphics[width=1.0\columnwidth]{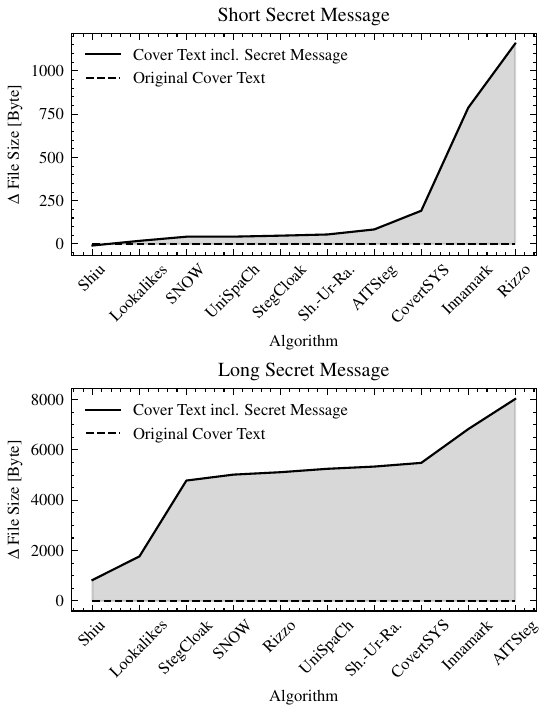}
		\caption{File size evaluation results (smaller absolute values are better).} \label{fig:file-size}
	\end{figure}

	In our testbed, we compared the differences in file size between the original cover text and the text including a secret message. Although these benchmark criteria are closely related to the previous analysis of the number of characters, the results differ due to varying character storage requirements. For example, Shazzad-Ur-Rahman et al.'s~\cite{ShazzadUrRahman.2021, ShazzadUrRahman.2023} method replaces a classical small Latin letter ``g'' (U+0067) with the visually similar-looking mathematical alphanumeric symbol U+1D5C0 as part of the letterlike symbols in the Unicode standard~\cite{TheUnicodeConsortium.2025}. Whereas the first needs one byte of storage space in the UTF-8 encoding, the second needs four bytes. Fig.~\ref{fig:file-size} illustrates the mean absolute differences based on the two runs on our 1\,000\,000-article dataset. The negative value for short secret messages from the Shiu et al.\ algorithm indicates a decreased file size. This can be attributed to replacing whitespace characters with newlines, as seen in the results for the number of characters. Our proposed Innamark technique is just in the lower third in direct comparison because it is one of the few algorithms that embeds the secret message multiple times in the cover text for increased robustness.
	
	\subsubsection{Caret Navigation} The last imperceptibility metric relates to suspicious behavior when navigating through a digital text document with the arrow keys on a computer keyboard. The \emph{caret} is the blinking pipe in a text field that indicates the cursor position. If, for example, a document contains zero-width characters and the caret arrives at such a character when arrow keys are used for navigation, the caret moves one zero-width character further but visually stays at the same position, raising suspicion. Similar abnormalities occur when multiple small whitespace characters are used.
	
	In our testbed, we manually checked various texts with secret messages. We moved the caret from the start to the end to identify any unusual behavior. In cases of hanging caret positions or identifying unexpected characters, we classified the algorithm as visible because the caret attack revealed the secret message's position. Table~\ref{tab:caret-navigation} shows an overview of the results with a justification for each algorithm as to why it is noticeable by users or not. It was found that only three algorithms (Rizzo et al., Lookalikes, and Innamark) are invisible against caret navigations.
	
	\begin{table}[htb]
		\centering
		\caption{Caret navigation attack evaluation.}\label{tab:caret-navigation}
		\setlength{\tabcolsep}{3pt}
		\begin{tabular}{p{55pt}cp{140pt}}
			\hline
			Name & Invisible & Reason\\
			\hline
	        SNOW & \ding{55} & Adds multiple whitespace characters at the end of the text.\\
			UniSpaCh & \ding{55} & Adds small spaces between words, sentences, and paragraphs.\\
			AITSteg & \ding{55} & Adds zero-width characters at the beginning of the text.\\
			Shiu et al. & \ding{55} & Adds whitespace characters at specific positions to encode 0 or 1.\\
			Rizzo et al. & \ding{51} & Only replaces characters without adding additional ones.\\
			StegCloak & \ding{55} & Adds zero-width characters in one position in the text.\\
			Lookalikes & \ding{51} & Only replaces letters without adding additional characters.\\
			CovertSYS & \ding{55} & Adds zero-width characters at the end of the text.\\
			Shazzad-Ur-Rahman et al. & \ding{55} & Replaces letters and single whitespace characters with combinations of multiple whitespace characters.\\
			Innamark & \ding{51} & Only replaces whitespace characters without adding additional characters.\\
			\hline
		\end{tabular}
	\end{table}

	\subsection{Robustness}
	\label{sec:robustness}
	The last primary benchmark criterion relates to the persistence of a secret message inside the text. We focus on the following two types of robustness checks:
	\begin{enumerate}
		\item modification robustness (insertion, replacement, deletion);
		\item usage robustness (retyping, formatting, file type and application change via copy and paste).
	\end{enumerate}
	
	\subsubsection{Modification Robustness}
\begin{figure}[htb]
		\centering
		\includegraphics[width=1.0\columnwidth]{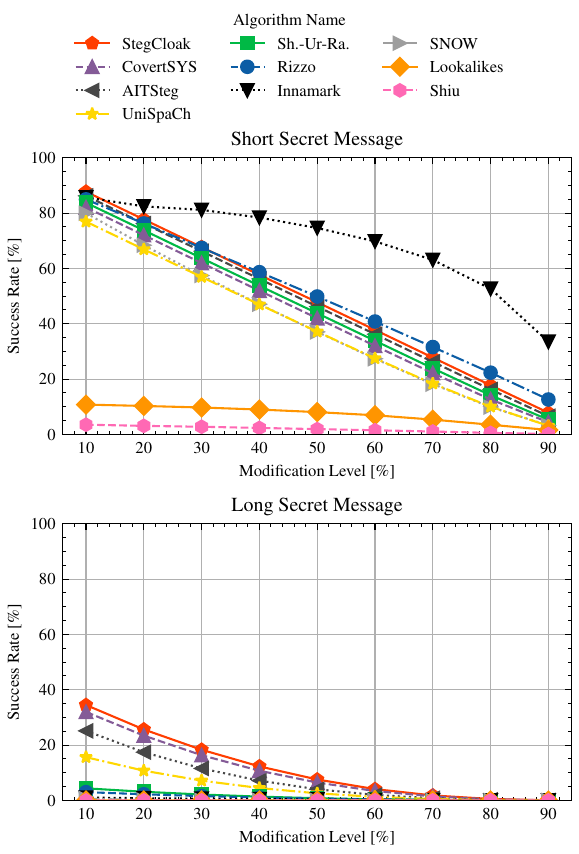}
		\caption{Modification robustness (higher values are better).} \label{fig:modification-robustness}
	\end{figure}

	Modifications to a file that may compromise the secret message, such as text insertions or deletions, are also known as tampering attacks in the information-hiding literature~\cite{Ahvanooey.2018b}. We have analyzed replacements, each consisting of a deletion followed by an insertion, for all implemented algorithms in batch runs for both the short and long secret messages on the 1\,000\,000-article dataset. For each article, we embedded the secret message and then replaced a block of 10\% of the length of the original cover text with just one letter. The starting position of the replacement was randomly chosen using seeding for reproducibility. A success rate was calculated, indicating the percentage of articles for which a secret message was successfully extracted after modification. Next, the process was repeated with replacements of 20\% up to 90\% in steps of 10 percentage points. This resulted in a total of 180\,000\,000 processed articles for the test (10 algorithms $\times $ 9 modification percentages $\times$ 2 execution runs for short and long secret message $\times$ 1\,000\,000 articles). The results are presented in Fig.~\ref{fig:modification-robustness}.
	
	For short secret messages, our proposed Innamark technique proved to be the most robust method in our testbed, even at high modification levels. Lookalikes and Shiu et al.\ had a particularly low success rate, but the other algorithms show a similar linear robustness trend.
	For long secret messages, all tested techniques are roughly similar, with StegCloak showing the best results because the algorithm embeds the secret message at a single position.
	
	\subsubsection{Usage Robustness}
	The usage robustness considered several aspects of how users work with text. We tested various usage scenarios classified as attacks on the information-hiding schemes on the full testbed of all implemented algorithms. The results are summarized in Table~\ref{tab:usage-robustness} and explained in the following.
	
	First, we analyzed a text reproduction or retyping attack~\cite{Jalil.2009,Ahvanooey.2018b}, in which users manually type the text with an embedded secret message in a new document. None of the algorithms could withstand this attack since they all use specific Unicode symbols that are lost on retyping.
	
	Second, we applied different formatting changes and checked whether the secret message could be extracted afterward~\cite{Ahvanooey.2022,Ahvanooey.2018b}. We embedded a secret message inside a cover, made the text bold, and changed the font, color, and size. The secret message could be successfully extracted in all cases because all algorithms work with Unicode symbols that are not affected by such styling attacks. Nevertheless, changing the font can affect the perceptibility of the secret messages if the algorithm uses specific characters unsupported by the font family.
	
	Third, and in accordance with the whitespace evaluation mentioned in Section~\ref{sec:notations}, we analyzed the robustness of a cover text with an embedded secret message when used in different applications and file formats. Simple copy and paste operations form ``one of the most common attacks in that the malicious users copy the whole of text and paste into their own files''~\cite[p.~7]{Ahvanooey.2018b}. Following~\cite{ShazzadUrRahman.2021}, we extended the set of business-related targets to social media applications, namely, WhatsApp, Facebook Messenger, and X (formerly Twitter).
	
	In our testbed, we applied each algorithm to a \emph{Lorem ipsum} dummy text to hide a secret message in the cover text. We copied each result into the relevant application and checked whether the secret message could still be extracted after copying it back to our testbed GUI. In Table~\ref{tab:usage-robustness}, ``\ding{51}'' indicates that the secret message could fully be extracted, whereas ``\ding{55}'' indicates a corrupted output. Edge cases are depicted as ``(\ding{51}),'' like CovertSYS in a .docx document, in which additional characters were shown in the extracted result but the original secret message could still be recognized. Only our proposed Innamark technique worked partially in the PDF format, depending on the PDF viewer used. For example, some whitespace characters were replaced with a standard U+0020 space when copying the content from Adobe Acrobat Reader, whereas they remained the same with PDF24 Reader. In such cases, ``(\ding{51})'' is shown in Table~\ref{tab:usage-robustness} because the respective whitespace characters remain in the original PDF file but the robustness depends on the PDF viewer used.
	
	\begin{table}[htb]
		\centering
		\caption{Usage robustness.}\label{tab:usage-robustness}
		\setlength{\tabcolsep}{3pt}
		\begin{tabular}{p{75pt}|cc|ccc|ccccc}
			\hline
			Algorithm & \rotatebox{90}{Retyping} & \rotatebox{90}{Formatting} & \rotatebox{90}{.txt} & \rotatebox{90}{.docx} & \rotatebox{90}{.pdf} & \rotatebox{90}{Mail} & \rotatebox{90}{Teams} & \rotatebox{90}{WhatsApp} & \rotatebox{90}{Facebook Msg. } & \rotatebox{90}{X/Twitter}\\
			\hline
			SNOW                        & \ding{55} & \ding{51} & \ding{51} & \ding{51} & \ding{55} & \ding{55} & \ding{55} & \ding{55} & \ding{55} & \ding{55} \\
			UniSpaCh                    & \ding{55} & \ding{51} & \ding{51} & \ding{51} & \ding{55} & \ding{51} & \ding{51} & \ding{51} & \ding{51} & \ding{51} \\
			AITSteg                     & \ding{55} & \ding{51} & \ding{51} & \ding{55} & \ding{55} & \ding{55} & \ding{51} & \ding{51} & \ding{51} & \ding{55} \\
			Shiu et al.                 & \ding{55} & \ding{51} & \ding{51} & \ding{51} & \ding{55} & \ding{51} & \ding{51} & \ding{51} & \ding{51} & (\ding{51}) \\
			Rizzo et al.                & \ding{55} & \ding{51} & \ding{51} & \ding{55} & \ding{55} & \ding{55} & (\ding{51}) & \ding{51} & \ding{51} & \ding{55} \\
			StegCloak                   & \ding{55} & \ding{51} & \ding{51} & \ding{51} & \ding{55} & \ding{51} & \ding{55} & \ding{51} & \ding{51} & \ding{51} \\
			Lookalikes                  & \ding{55} & \ding{51} & \ding{51} & \ding{51} & \ding{55} & \ding{51} & \ding{51} & \ding{51} & \ding{51} & \ding{51} \\
			CovertSYS                   & \ding{55} & \ding{51} & \ding{51} & (\ding{51}) & \ding{55} & \ding{55} & \ding{51} & \ding{51} & \ding{55} & \ding{55} \\
			Shazzad-Ur-Rahman et al.    & \ding{55} & \ding{51} & \ding{51} & \ding{51} & \ding{55} & \ding{51} & \ding{55} & \ding{51} & \ding{51} & \ding{51} \\
			Innamark                    & \ding{55} & \ding{51} & \ding{51} & \ding{51} & (\ding{51}) & \ding{51} & \ding{51} & \ding{51} & \ding{51} & \ding{51} \\
			\hline
		\end{tabular}
	\end{table}
	
	The results show that some applications remove specific characters like the four-per-em space (U+2005), which do not work in file types like .docx and PDF or in emails (see Table~\ref{tab:whitespace-evaluation}) but are used by algorithms like Rizzo et al.~\cite{Rizzo.2019}. Further, the tested messenger software and social media networks often remove trailing whitespace characters from messages, which is why SNOW and CovertSYS encountered problems. Only our proposed Innamark technique was robust in all tested applications and file types.
	
	\section{Discussion}
	\label{sec:discussion}
	This section discusses our proposed Innamark scheme on the basis of the experimental evaluation results, considering its limitations and future research directions. We evaluated and compared it against a testbed of ten algorithms from the literature, conducting tests on a dataset of 1\,000\,000 articles. To our knowledge, the reported Innamark algorithm is the first method that can hide a secret message inside a cover text without increasing the number of characters or being noticed by humans while ensuring robustness to replacing portions of text and copying it between applications.
	
	\begin{figure}[htb]
		\centering
		\includegraphics[width=0.7\columnwidth]{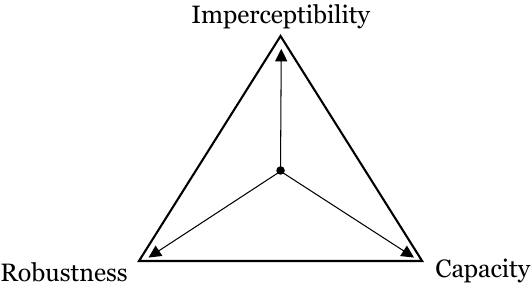}
		\caption{Requirement trade-off based on~\cite{Li.2021}.} \label{fig:trade-off-triangle}
	\end{figure}
	\begin{figure*}[thb]
		\centering
		\includegraphics[width=1.0\textwidth]{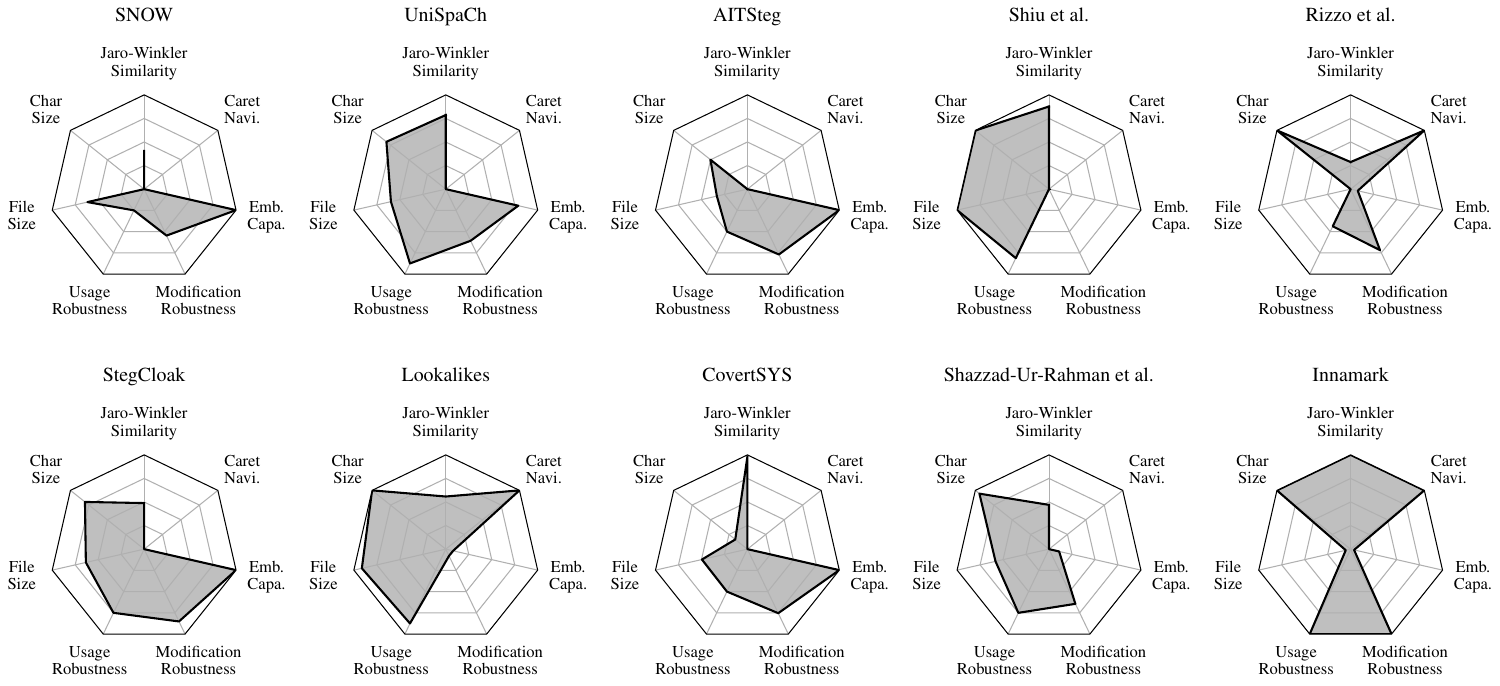}
		\caption{Summarizing evaluation comparison inspired by~\cite{Knochel.2024}.} \label{fig:radar-summary}
	\end{figure*}
	
	In general, it should be emphasized that there is no one-fits-all information-hiding scheme with high robustness, embedding capacity, and imperceptibility. Information-hiding techniques often strive for high embedding capacities, which often conflict with imperceptibility criteria since imperceptibility decreases if the embedding capacity increases~\cite{Yang.2019}. This can be illustrated with a \emph{trade-off triangle}~\cite{Li.2021}, shown in Fig.~\ref{fig:trade-off-triangle}. Therefore, researchers and practitioners need to select an appropriate algorithm for their use case under the consideration of boundary conditions and application scenarios. Following~\cite{Knochel.2024}, we provide an overview in Fig.~\ref{fig:radar-summary} based on the evaluation results to support the decision-making process. 
	
	An unbounded algorithm like AITSteg, CovertSYS, StegCloak, or SNOW should be selected if a high embedding capacity is essential. Our Innamark algorithm is a favorable choice if data are often transferred between different applications and both robustness and imperceptibility are important.
	
	\subsection{Limitations and Future Work}
	Nevertheless, our approach has limitations that need to be discussed in future work.
	
	First, Innamark's small embedding capacity is a weakness. The structure of an InnamarkTag, as shown in Fig.~\ref{fig:innamarktag-structure}, is designed to enable optional compression that can help to increase the capacity. Work is in progress to increase the capacity further by using compression libraries or developing hybrid approaches that use non-printable characters. Furthermore, the impact of using different InnamarkTag options needs to be analyzed to determine to what extent the error correction improves robustness or how hashing affects the embedding capacity.
	
	Second, the proposed solution is based on and tested on the Unicode standard and the UTF-8 scheme. Future research should consider the influence of other encoding schemes like UTF-16 and ISO-standardized Latin-1 (ISO 8859-1), and the potential impacts on the process of recoding to a small scheme like ASCII.
	
	Third, the algorithm is robust to the actions of typical users who do not recognize documents with hidden secret messages. However, people familiar with the strategy can use smart attacks to apply targeted destruction of the secret message, for example, the random replacement of $\mathcal{A}_+$. This can be eliminated by using a different random subset of $\mathcal{A}_+$ for every embedding operation. With the help of the modular InnamarkTag structure, future research can use smart analysis in encoding to increase the noise tolerance and allow the extraction and restoration of broken secret messages.
	
	Fourth, printing and rescanning attacks using \ac{OCR} may destroy secret messages. However, most spaces have slightly different widths that are not recognizable by human eyes, so machines may restore the original whitespace characters from a scan if configured correctly. This strongly depends on the font used, the applied \ac{OCR} technique, and physical conditions like the scan quality.
	
	Fifth, accessibility concerns may arise if a text with a secret message is read by screen readers or automatically translated into another language. Future research is needed to analyze how those accessibility tools handle specific Unicode characters. This emphasizes the need for a deeper analysis of related text processing tools, such as machine learning-based \ac{NLP} units, minifiers, compressors, or deobfuscation tools.

	\section{Conclusion}
	\label{sec:conclusion}
	We have designed and implemented Innamark, a blind and invisible information-hiding technique that can embed byte-encoded sequences inside a cover text. Although several solutions for digital text watermarking and steganography have been published in recent years, existing approaches change the semantics or style of the cover text, increase the number of characters, or lack robustness against the output being copied into different applications. By encoding and mapping a secret message into our embedding alphabet of five Unicode whitespace characters, we can embed the information in the cover text by substituting all whitespace characters. The specified structure of our InnamarkTag has been designed to enable additional functionalities like compression, encryption, hashing, and error correction. The experimental evaluation shows strengths in imperceptibility and robustness, with limitations in embedding capacity based on a direct benchmark comparison with ten algorithms. Our method can help \ac{LLM} operators fulfill regulations like the European Union's AI Act~\cite{EuropeanCommission.2024} and can assist businesses in securing sensitive data before it is shared with external parties, especially if they are concerned about robustness to copying between applications. Work is in progress to increase the algorithm's embedding capacity and enable the restoration of secret messages broken by text alterations.
	
	\section*{Appendix}
	\subsection*{Evaluation Data}
	To evaluate, compare, and benchmark our proposed Innamark method against existing solutions, we performed two batch runs with each algorithm. The first used the English example name ``John'' as a short secret message. The second used the following 455-character \emph{Lorem ipsum} dummy text as a long secret message:
	``Lorem ipsum dolor sit amet, consectetur adipiscing elit, sed do eiusmod tempor incididunt ut labore et dolore magna aliqua. Ut enim ad minim veniam, quis nostrud exercitation ullamco laboris nisi ut aliquip ex ea commodo consequat. Duis aute irure dolor in reprehenderit in voluptate velit esse cillum dolore eu fugiat nulla pariatur. Excepteur sint occaecat cupidatat non proident, sunt in culpa qui officia deserunt mollit anim id est laborum.''
	
	The set of cover texts consists of 1\,000\,000 random English Wikipedia articles, chosen because they are publicly available texts that have different lengths and structures and cover a wide range of domains. We used the cleaned article version from Hugging Face based on the dump from the Wikimedia Foundation~\cite{WikimediaFoundation.2023}. To reproduce our randomized selection, the articles’ IDs are available from the corresponding author upon request and can be mapped back to the original texts and URLs.

	\footnotesize
	\section*{Acknowledgements}
	This work was supported by the Cluster of Excellence Cognitive Internet Technologies CCIT and the Center of Excellence Logistics and IT, which are funded by the Fraunhofer-Gesellschaft.
	
	\normalsize
	\bibliographystyle{IEEEtran}
	\bibliography{literature}
	
\end{document}